\documentclass[twocolumn,english,prb]{revtex4}
\usepackage[T1]{fontenc}
\usepackage[latin9]{inputenc}
\usepackage{amsmath}
\usepackage{amssymb}
\usepackage{graphicx}
\usepackage{wasysym}

\makeatletter
\@ifundefined{textcolor}{}
{%
 \definecolor{BLACK}{gray}{0}
 \definecolor{WHITE}{gray}{1}
 \definecolor{RED}{rgb}{1,0,0}
 \definecolor{GREEN}{rgb}{0,1,0}
 \definecolor{BLUE}{rgb}{0,0,1}
 \definecolor{CYAN}{cmyk}{1,0,0,0}
 \definecolor{MAGENTA}{cmyk}{0,1,0,0}
 \definecolor{YELLOW}{cmyk}{0,0,1,0}
}

\makeatother

\usepackage{babel}
\begin{document}

\title{Transport signatures in topological systems coupled to AC fields}

\author{Leonard Ruocco and Álvaro Gómez-León}

\affiliation{Department of Physics and Astronomy and Pacific Institute of Theoretical
Physics\\
University of British Columbia, 6224 Agricultural Rd., Vancouver,
B.C., V6T 1Z1, Canada.}
\email{agomez@phas.ubc.ca}

\selectlanguage{english}%

\date{\today}
\begin{abstract}
We study the transport properties of a topological system coupled
to an AC electric field by means of Floquet-Keldysh formalism. We
consider a semi-infinite chain of dimers coupled to a semi-infinite
metallic lead, and obtain the density of states and current when the
system is out of equilibrium. Our formalism is non-perturbative and
allows us to explore, in the thermodynamic limit, a wide range of
regimes for the AC field, arbitrary values of the coupling strength
to the metallic contact and corrections to the wide-band limit (WBL).
We find that hybridization with the contact can change the dimerization
phase, and that the current dependence on the field amplitude can
be used to discriminate between them. We also show the appearance
of side-bands and non-equilibrium zero-energy modes, characteristic
of Floquet systems. Our results directly apply to the stability of
non-equilibrium topological phases, when transport measurements are
used for their detection.
\end{abstract}
\maketitle

\paragraph*{Introduction:}

\begin{figure}
\includegraphics[scale=0.5]{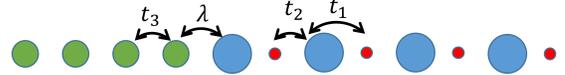}\caption{\label{fig:Schematic}Schematic figure of a dimers chain with sites
A/B(blue/red) coupled to a metallic contact (green). The hopping $t_{1,2}$
characterize the hop of electrons in the chain, and $t_{3}$ the hop
in the metallic contact. Finally, the two systems are coupled by $\lambda$,
which allows for the electrons to hop between the last A site of the
dimers chain and the metallic contact.}

\end{figure}

Systems with topological properties are of great interest due to their
unusual bulk/edge physics. In materials realizing these states of
matter, the bulk usually corresponds to an insulator while the edge
contains localized modes with interesting transport properties\cite{Bernevig2006a,TI-Zhang1,Hasan2010}.
While the study of topological systems with weak interactions has
led to a very complete understanding of their bulk physics during
the last years\cite{ClassificationTI-Schnyder2008}, their external
control and detection is still a very active field of research, with
many paradigms still to be understood\cite{Majorana-Detection-Cirac,Rokhinson2012,Nature-Majorana-2014,Eisert2015}.
A very interesting proposal is to induce a topological phase in an
initially trivial system by means of an external driving. Several
approaches have been discussed in the literature, such as shaken optical
lattices\cite{ShakenOpLatt1} or photo-induced states\cite{Lindner2010,Kitagawa2011,AlviDimers,Alvi-Anomalous,Bello2015-DimersAc,Gedik2013,Delplace2013,Grushin2014},
but most of them rely on the same principle. In this work we study
the transport signatures of an AC driven semi-infinite chain of dimers,
when it is connected to a metallic reservoir (see Fig.\ref{fig:Schematic}).
The dimers chain is a very interesting system due to its simple mathematical
description, its non-trivial topological properties\cite{Hatsugai,AlviDimers}
and its connection with graphene ribbons\cite{Delplace2011-Graphene}
and soliton physics\cite{SSH-1979,SSH-1980,SSH-Review1988}. Furthermore,
their application in molecular electronics has been previously studied
in the absence of AC fields\cite{SolitonSwitch,MolecularSwitch}.
In this work we study the edge and bulk properties of the non-equilibrium
topological phase of a dimers chain. We obtain the surface Green's
functions for the case of a semi-infinite chain in Keldysh formalism\cite{Brandes1997-Keldysh,Wu2008-Keldysh-Dot,Aoki2014-DMFT,Boltzmann-Keldysh,Pollito}
and study the current through the system as a function of the parameters
of the external field and coupling strength to the metallic contact.
The combination of surface Green's functions\cite{Book-MolecularTransport2009,Velev2004}
with the Floquet-Keldysh formalism allows us to obtain expressions
for the transport in very interesting regimes, which do not rely on
perturbative expansions or master equations, which can fail in some
cases\cite{ViolationOnsager-Seja2016} and exclude memory effects/backscattering.
As the dimer chain is semi-infinite, we obtain the Green's functions
for the edge modes in the thermodynamic limit. Our results discuss
the fate of the topological properties once the system is coupled
to a measurement apparatus, and finite frequency corrections to the
well known Magnus expansion in the high frequency regime\cite{Blanes2009-Magnus,Eckardt2015,Alessio2015}.

\paragraph*{Model:}

We consider the following Hamiltonian for the time dependent system:
\begin{equation}
H\left(t\right)=H_{D}+H_{C}+H_{T}+H_{AC}\left(t\right)\label{eq:Hamiltonian}
\end{equation}
where the different terms correspond to the dimers chain, metallic
contact, tunneling and coupling to the AC field, respectively. Concretely,
each term is given by:
\begin{eqnarray}
H_{D} & = & -\sum_{\alpha,\beta}\sum_{\langle i,j\rangle}t_{i,j}^{\alpha,\beta}d_{i,\alpha}^{\dagger}d_{j,\beta}\\
H_{C} & = & \mu_{L}\sum_{i}c_{i}^{\dagger}c_{i}-t_{3}\sum_{\langle i,j\rangle}c_{i}^{\dagger}c_{j}\\
H_{T} & = & \sum_{j}\sum_{i,\sigma}\left(\lambda_{j;i,\sigma}^{\ast}c_{j}^{\dagger}d_{i,\sigma}+\lambda_{j;i,\sigma}d_{i,\sigma}^{\dagger}c_{j}\right)\\
H_{AC}\left(t\right) & = & qV_{0}\left(t\right)\sum_{i}x_{i,\alpha}d_{i,\alpha}^{\dagger}d_{i,\alpha}
\end{eqnarray}
where $c_{i}^{\dagger}$ creates a spinless fermion at site $i$ of
the contact, $d_{i,\sigma}^{\dagger}$ creates a spinless fermion
at site $i$ and sub-lattice $\sigma$ of the dimers chain, $\mu_{L}$
is the chemical potential in the contact, $t_{i,j}^{\alpha,\beta}$
is the nearest neighbors hopping in the dimers chain, $t_{3}$ the
nearest neighbors hopping in the metallic contact and $\lambda_{j;i,\sigma}$
is the tunneling connecting the two systems (we choose the relevant
case of nearest neighbors tunneling, although more general situations
are possible). As the system under consideration corresponds to a
bipartite lattice, it will simplify some expressions to rename the
$\alpha=A,B$ index to $\alpha=+,-$; then both conventions are considered
indistinguishable. The coupling to the AC field can be written in
different ways, and here we have considered the dipolar coupling of
the electric field to the local charge density of the system, being
$V_{0}\left(t\right)$ the voltage, $q$ the electric charge and $x_{i,\alpha}$
the position of site $i$ in sub-lattice $\alpha$\cite{Benito-AC-Dimers,Monica1}.
Another standard method to introduce the driving field is to consider
the temporal gauge, where the scalar potential $\phi$ vanishes and
the vector potential $A\left(t\right)$ is time dependent\cite{Aoki2014-DMFT}.
Then, by means of the minimal coupling $k\rightarrow k+qA\left(t\right)$,
one obtains the time dependent Hamiltonian. Importantly, $H\left(t\right)$
can be transformed into the Hamiltonian in the temporal gauge by going
to the interaction picture $\tilde{H}\left(t\right)=U\left(t\right)H\left(t\right)U\left(t\right)^{\dagger}-iU\left(t\right)\dot{U}\left(t\right)^{\dagger}$,
where $U\left(t\right)=\exp\left\{ i\int H_{AC}\left(t\right)dt\right\} $\cite{Alessio2015}.
Therefore, both cases are equivalent and we can choose any of them
without loss of generality -each case corresponds to a different gauge
choice. Note that these different couplings to the driving field cover
a wide range of physical realizations, e.g., light irradiation to
the dimers chain, a time dependent gate voltage, or the shake of an
optical lattice. Although the effect of electron-electron interactions
is out of the scope of this work, they can be included in the self-energies.
In the presence of interactions, novel topological features could
be obtained when they are strong enough\cite{Gomez-Leon2016}, and
their interplay with the AC field and their detection by transport
measurements would be interesting for future works.

\begin{figure}
\includegraphics[scale=0.8]{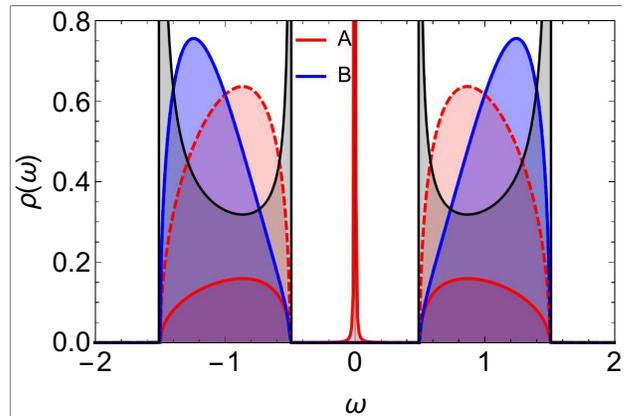}

\caption{\label{fig:DOS-EdgeDimers}Bulk (black) and surface (red and blue)
DOS for the isolated dimers chain. The solid lines correspond to the
topological phase ($t_{1}/t_{2}=0.5$) and the dashed lines to the
trivial one ($t_{1}/t_{2}=2$); for the B site the S-DOS overlap,
as the only difference between the trivial and the topological is
in the zero energy mode at A. The bulk DOS shows the gap between the
conduction and valence band, as well as the Van Hove singularities
expected from the flat dispersion at the top/bottom of the bands.}
\end{figure}

We first investigate the undriven bulk and surface Green's functions
in the dimers chain. The advantage of the surface Green's functions
for the case of semi-infinite systems is double fold, on the one hand
one can obtain exact analytical expressions for the edge modes when
the system is infinitely large in one direction, but has a boundary
in the other one (thermodynamic limit in which the domain walls are
infinitely far, and do not interact); on the other hand, surface Green's
functions are essential for the calculation of the current. In order
to obtain the surface Green's functions one just needs to make use
of Dyson's equation:
\begin{equation}
\hat{G}=\hat{g}+\hat{g}\cdot\hat{\Sigma}\cdot\hat{G}\label{eq:Dyson1}
\end{equation}
where $\hat{g}$ corresponds to the matrix Green's function of a semi-infinite
chain and a single site initially decoupled, $\hat{\Sigma}$ to the
self-energy representing the coupling via tunneling of the single
site to the semi-infinite chain, and $\hat{G}$ to the total Green's
function to be determined (details in the Appendix). Noticing that
for a semi-infinite system, the unperturbed Green's function of the
chain and the perturbed one for the single site must be the same,
one obtains a quadratic equation, whose solution provides the surface
Green's function. From this expression one can obtain the surface-density
of states(S-DOS) using $\rho_{0}^{S}\left(\omega\right)=\pm\frac{1}{\pi}\underset{\epsilon\rightarrow0^{\pm}}{\lim}\Im\left\{ G_{0}\left(\omega\mp i\eta\right)\right\} $,
where $G_{0}\left(\omega\right)$ corresponds to the perturbed surface
Green's function obtained from Eq.\ref{eq:Dyson1}. The calculation
for both, the case of the linear and the dimers chain is equivalent,
and the only difference is the increase in the matrix size due to
the sub-lattice degree of freedom. The previous result provides the
surface Green's function for the isolated, semi-infinite dimers chain;
in order to include the effect of hybridization with the metallic
contact we need to solve Dyson's equation again, with the self-energy
produced by the hopping to the linear chain (in this second case it
corresponds to a simple matrix inversion). In Fig.\ref{fig:DOS-EdgeDimers}
we plot the S-DOS at site $A$ (red) and $B$ (blue) of the dimers
chain, and a comparison with the bulk DOS (black) $\rho^{B}\left(\omega\right)$.
The bulk DOS is obtained form the Green's function of a dimers chain
with periodic boundary conditions.

As expected for $t_{1}/t_{2}<1$, the S-DOS at site $A$ shows a zero-energy
mode due to the topological nature of the system and the open boundary
conditions. This effect is well known and was first predicted in polyacetylene
chains\cite{SSH-1979}, where the phonon field exhibits a degenerate
ground state (two dimerization states) with solitonic excitations,
and the coupling of the electrons to the solitons induces pairs of
domain walls with localized electronic modes. In our model, the domain
walls are created by a change in the hopping parameter, which is equivalent
to a spatial modulation of the mass of the fermionic field. The topological
properties of this system in equilibrium have been widely discussed
in the literature\cite{Hatsugai}, and its non-equilibrium counterpart
is well understood for the case of an isolated chain\cite{AlviDimers,Delplace-QuantumWalk}.
In equilibrium and for nearest neighbors hopping the two different
topological phases can be characterized by a winding number $\nu_{1}=\left\{ 0,1\right\} $
depending on the ratio $t_{1}/t_{2}$. When $\nu_{1}=1$ the system
is said to be topological and displays localized edge modes as the
one seen in Fig.\ref{fig:DOS-EdgeDimers}. Out of equilibrium the
classification is more complicated, and the appearance of two unequivalent
gaps in the Floquet quasi-energy spectrum leads to a topological index
$\mathbb{Z}\times\mathbb{Z}$\cite{Chiral-QWalks,Delplace-QuantumWalk}.
In this work we will not calculate the topological invariants, but
rather we will focus on their experimental signatures in the DOS,
and on the differences in the transport properties between the topological
and the non-topological phases.

\begin{figure}
\includegraphics[scale=0.8]{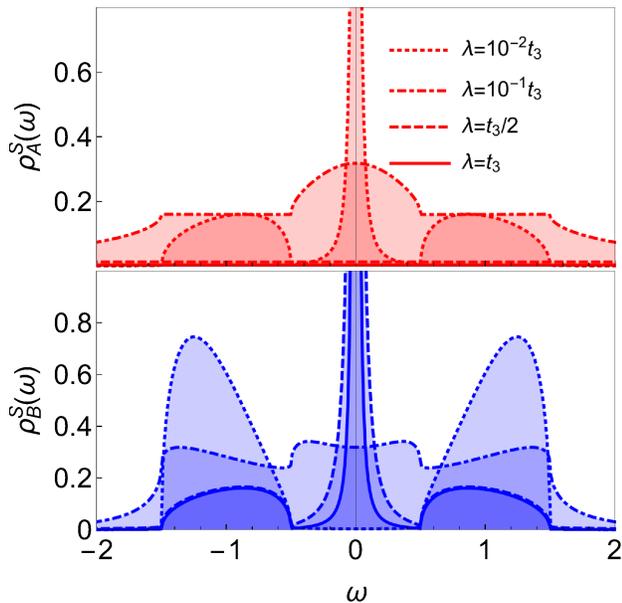}

\caption{\label{fig:Hybridization}S-DOS at the A(B) site for $t_{1}/t_{2}=0.5\left(2\right)$
(up/down, respectively) and different values of $\lambda$. The initially
topological phase becomes trivial when the A atom hybridizes with
the contact. In opposition, for the trivial phase a zero energy mode
emerges as $\lambda$ increases; this is because the hybridization
exchanges the dimerized phase ($t_{1}\leftrightarrow t_{2}$).}

\end{figure}

We now discuss the effect of hybridization with the metallic contact
for the case of nearest neighbors hopping $\lambda_{j;i,\sigma}=\lambda\delta_{i,j}\delta_{\sigma,A}$
(we choose this specific form for all calculations, but the generalization
to a larger number of neighbors is straightforward). Intuitively,
one would guess that if we start with our system in the topological
phase ($t_{1}/t_{2}<1$), as the isolated edge state directly couples
to the contact, it would be the one mostly affected. This is precisely
what happens, especially for $\lambda\ll t_{3}$, where we find that
the main effect in the S-DOS is the widening of the zero energy mode,
although it remains well defined up to $\lambda\sim0.1t_{3}$ (see
Fig.\ref{fig:Hybridization}). Larger values of $\lambda$ smear out
the zero energy mode until $\lambda\sim t_{3}$, where it merges with
the continuum and any signature of the zero energy mode disappears.
More counter-intuitive is the fact that the opposite process can also
happen, where the A atom of the trivial phase hybridizes with the
contact and B becomes the effective last site of the chain. This is
equivalent to changing the dimerization ground state, and transforms
the system into its topological phase, with the appearance of a zero
energy mode at site B. 

\begin{figure}
\includegraphics[scale=0.8]{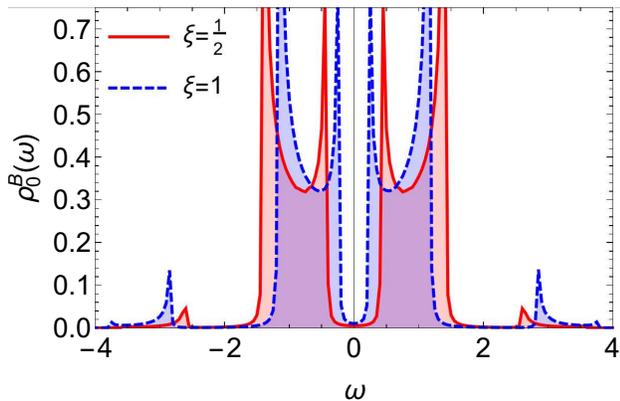}

\caption{\label{fig:DOS-AC-Dimers}Bulk time averaged DOS for different values
of the AC field amplitude $\xi$. The AC induces side-bands at $\omega=n\Omega$
and the field amplitude $\xi$ controls their width. Note that as
we increase $\xi$, the gap size decreases, and for $t_{2}\mathcal{J}_{0}\left(\xi\right)=t_{1}$
we obtain the DOS of a linear chain. An increase of $\xi$ increases
the spectral weight of the other side-bands. We have chosen $\Omega=4$
and $t_{1}=0.5$, in units of $t_{2}$.}
\end{figure}

Now that we have characterized our system properties in equilibrium
we discuss the effect of the AC field. The reason why we do not initially
consider the temporal gauge in Eq.\ref{eq:Hamiltonian} is based on
the absence of translational symmetry for a semi-infinite chain, however,
the transformation to the interaction picture will still help us to
encode the effect of the AC field in the hopping, and simplify the
calculations. The transformation leads to the following time dependent
Hamiltonian:
\begin{eqnarray}
\tilde{H}\left(t\right) & = & \tilde{H}_{D}\left(t\right)+H_{C}+\tilde{H}_{T}\left(t\right)\\
\tilde{H}_{D}\left(t\right) & = & -\sum_{i,j,\alpha,\beta}\tilde{t}_{i,j}^{\alpha,\beta}\left(t\right)d_{i,\alpha}^{\dagger}d_{j,\beta}\\
\tilde{H}_{T}\left(t\right) & = & \sum_{j,i,\sigma}\left[\tilde{\lambda}_{j;i,\sigma}^{\ast}\left(t\right)c_{j}^{\dagger}d_{i,\sigma}+\tilde{\lambda}_{j;i,\sigma}\left(t\right)d_{i,\sigma}^{\dagger}c_{j}\right]
\end{eqnarray}
where the time dependent terms are $\tilde{t}_{i,j}^{\alpha,\beta}\left(t\right)=t_{i,j}^{\alpha,\beta}e^{iq\left(x_{i,\alpha}-x_{j,\beta}\right)\int V_{0}\left(t\right)dt}$
and $\tilde{\lambda}_{j;i,\sigma}\left(t\right)=\lambda_{j;i,\sigma}e^{iqx_{i,\sigma}\int V_{0}\left(t\right)dt}$.
In the presence of a time dependent field, the Green's functions now
depend on $t$ and $t^{\prime}$ independently (do not confuse the
time coordinate with the hopping parameters $t_{1,2,3}$); however
the time periodicity of the Hamiltonain ensures that $\hat{G}\left(t+T,t^{\prime}+T\right)=\hat{G}\left(t,t^{\prime}\right)$.
This symmetry can be used in our advantage if we consider Wigner coordinates
$t_{+}=\left(t+t^{\prime}\right)/2$ and $t_{-}=t-t^{\prime}$, which
imply that $\hat{G}\left(t_{-},t_{+}+T\right)=\hat{G}\left(t_{-},t_{+}\right)$.
We define the Floquet-Green's function as\cite{Aoki2014-DMFT}:
\begin{equation}
\hat{G}_{mn}\left(\omega\right)\equiv\frac{1}{T}\int_{0}^{T}dt_{+}e^{i\left(m-n\right)\Omega t_{+}}\hat{G}\left(\omega+\frac{m+n}{2}\Omega,t_{+}\right)
\end{equation}
with $\hat{G}\left(\omega,t_{+}\right)=\int_{-\infty}^{\infty}e^{i\omega t_{-}}\hat{G}\left(t_{+},t_{-}\right)dt_{-}$.
The advantage of this representation is two fold: it separates long
time and short time dynamics, which allows for a simple physical interpretation,
and transforms time convolutions into matrix products. Then one has
the following simple form for the Dyson's equation:
\begin{equation}
\hat{G}_{mn}\left(\omega\right)=\hat{g}_{mn}\left(\omega\right)+\sum_{m^{\prime}n^{\prime}}\hat{g}_{mm^{\prime}}\left(\omega\right)\hat{\Sigma}_{m^{\prime}n^{\prime}}\left(\omega\right)\hat{G}_{n^{\prime}n}\left(\omega\right)\label{eq:Dyson}
\end{equation}
which highly simplifies the calculations of the time dependent surface
Green's functions. We now study the effect of the AC field on both,
the bulk and the surface Floquet-Green's functions of the dimers chain.
For the explicit calculations we fix $V_{0}\left(t\right)=V_{0}\cos\left(\Omega t\right)$,
although the formalism allows for more general AC fields. The calculation
of the bulk Green's functions is done using the equation of motion
technique (details in the Appendix), while for the case of the surface
Green's functions we consider Eq.\ref{eq:Dyson} to find the unperturbed
surface Green's function of the dimers chain, and then we add the
effect of the metallic contact via the self-energy. For the bulk Green's
function $G_{k}^{\alpha,\beta}\left(t,t^{\prime}\right)=-i\theta\left(t-t^{\prime}\right)\langle\left\{ d_{k,\alpha}\left(t\right),d_{k,\beta}^{\dagger}\left(t^{\prime}\right)\right\} \rangle$,
where $k$ is a good quantum number due to the periodic boundary condition,
we find the next hierarchy of equations in Wigner coordinates (remember
that $\alpha,\beta=\pm$ refers to the sublattice degree of freedom):
\begin{multline}
\left(n\frac{\Omega}{2}+\omega\right)G_{k;n}^{\alpha,\beta}\left(\omega\right)=\delta_{n,0}\delta_{\alpha,\beta}-t_{1}G_{k;n}^{-\alpha,\beta}\left(\omega\right)\\
-t_{2}\sum_{l}\mathcal{J}_{\alpha l}\left(\xi\right)e^{i\alpha k}G_{k;n+l}^{-\alpha,\beta}\left(\omega+l\frac{\Omega}{2}\right)\label{eq:Bulk-dimer}
\end{multline}
where $\mathcal{J}_{n}\left(\xi\right)$ is the $n$-th Bessel function
of the first kind and $\xi=qV_{0}/\Omega$. Note that the terms with
$l\neq0$ correspond to photon assisted tunneling, and lead to the
appearance of side-bands. Importantly, the coupling to different side-bands
rapidly decreases if $\Omega$ is the dominant energy scale, as contributions
to the DOS from processes absorbing/emitting a photon with $n\Omega$
energy are proportional to $\mathcal{J}_{n}^{2}\left(\xi\right)/\Omega^{2}$.
In this work we focus on high/intermediate frequency regimes $\Omega\apprge t_{1,2}$,
as this configuration shows interesting properties\cite{AlviDimers};
however our calculation includes arbitrary photon transitions until
we find numerical convergence, and can be used to study lower frequency
regimes. If to lowest approximation, we neglect transitions to different
side-bands, we find the usual result, viz. a renormalization of the
hopping $t_{2}$ by $\mathcal{J}_{0}\left(\xi\right)$. This means
that the DOS gets squeezed as the field amplitude increases, and for
a zero of the Bessel function, the system displays flat bands. When
we include higher order photon processes the renormalization is accompanied
by the appearance of side-bands at multiples $\omega=n\Omega$, which
can be observed in the time averaged DOS (Fig.\ref{fig:DOS-AC-Dimers}
shows the appearance of the first side-bands around $\omega=\pm2.5$),
and create extra transport channels which result in the Floquet sum
rule for the conductivity\cite{FloquetSumRule}.

\begin{figure}
\includegraphics[scale=0.8]{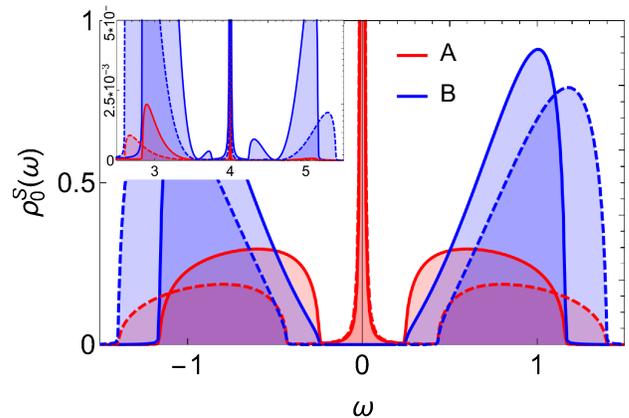}

\caption{\label{fig:SurfaceDOS-AC-Dimers}Time averaged S-DOS for the AC driven
dimers chain and field strength $\xi=\left\{ 0.5,1\right\} $(dashed
and solid, respectively). Note that the driving induces side-bands
centered at frequency multiples $\omega=n\Omega$ which display localized
modes as well (inset), but in contrast with the undriven case, the
$B$ site also has a finite spectral weight. For this plot we have
chosen $t_{1}=0.5$, $\lambda=0.01$ and $\Omega=4$ (in units of
$t_{2}$).}

\end{figure}

We now discuss the surface Green's functions in the presence of driving.
As we previously discussed, they are calculated using Eq.\ref{eq:Dyson}
and the Floquet form of the time-dependent self energy $\Sigma\left(t,t^{\prime}\right)$,
which couples the $B$ site of an isolated dimer to the $A$ atom
of the semi-infinite chain. Then, one can include the effect of the
metallic contact by direct matrix inversion of the corresponding Dyson's
equation. In Fig.\ref{fig:SurfaceDOS-AC-Dimers} we show the presence
of a zero-energy mode when $t_{1}/t_{2}\mathcal{J}_{0}\left(\xi\right)<1$,
meaning that the coupling to the AC field does not destroy the initial
topological phase due to the interaction between side-bands. Furthermore,
we also find ``zero-energy modes'' at multiples of $\omega=n\Omega$,
not only for the $A$ site, but for the $B$ site as well (only at
$n\neq0$, see Fig.\ref{fig:SurfaceDOS-AC-Dimers}). This feature
indicates that the edge states and topological phases of non-equilibrium
systems are, in general, different to those in undriven systems\cite{Chiral-QWalks,Benito2014,Delplace-QuantumWalk}.
The fact that the zero-energy modes at the $B$ site do not appear
in the $n=0$ side-band indicates that they occur dynamically, and
therefore possess an intrinsic time dependence. However, the oscillations
between the $A$ and $B$ site of the last dimer still correspond
to a localized zero-energy mode in the last dimer of the chain. The
occupation of both sub-lattices in the presence of driving can be
understood in terms of the extra symmetries present in periodically
driven systems\cite{Chiral-QWalks}. Importantly in our calculation,
the domain walls are infinitely far and there is no hybridization
between them; the effect corresponds to a purely dynamical one that
persists in the thermodynamic limit. 
The previous results show that the AC field produces several effects:
1) the bandwidth is renormalized by the field intensity, and this
can produce metal-insulator transitions due to the appearance/disappearance
of gaps; 2) the side-band structure implies the appearance of new
transport channels; and 3) it can drive topological phase transitions
with properties different to those of systems in thermal equilibrium.
For the characterization of these changes, we calculate the current
when the system is coupled to a metallic contact. We also describe
how hybridization between the two systems $\lambda$ influences the
current, and how the presence of zero-energy modes is captured in
the current profile. For this, we calculate the current operator\cite{Jauho1993}:
\begin{equation}
J\left(t\right)=2q\sum_{i,j,\sigma}\Re\left\{ \lambda_{j;\sigma,i}\left(t\right)G_{j;i,\sigma}^{<}\left(t,t\right)\right\} 
\end{equation}
being $G_{j;i,\sigma}^{<}\left(t,t^{\prime}\right)=i\langle d_{i,\sigma}^{\dagger}\left(t^{\prime}\right)c_{j}\left(t\right)\rangle$
the mixed lesser Green's function, and $\lambda_{j;\sigma,i}\left(t\right)$
the tunneling between the contact and the dimers chain\footnote{We assume $\lambda_{j;\sigma,i}\left(t\right)$ time independent at
the end of the calculation, and by choosing the contact point as $x_{A}=0$
we can write its interaction picture representation as time independent
as well.}. We can separate the mixed Green's function using the Langreth rules:
\begin{eqnarray}
J\left(t\right) & = & 2q\int_{-\infty}^{\infty}\Re\left\{ G_{0,A}^{r}\left(t,t_{1}\right)\Sigma_{0,A}^{<}\left(t_{1},t\right)\right.\nonumber \\
 &  & \left.+G_{0,A}^{<}\left(t,t_{1}\right)\Sigma_{0,A}^{a}\left(t_{1},t\right)\right\} dt_{1}
\end{eqnarray}
where we have fixed $\lambda_{j;i,\sigma}=\lambda\delta_{i,j}\delta_{\sigma,A}$,
$\Sigma_{0,A}^{a}\left(t_{1},t\right)=\left|\lambda\right|^{2}\mathfrak{g}_{0}^{a}\left(t_{1},t\right)$,
$\Sigma_{0,A}^{<}\left(t_{1},t\right)=\left|\lambda\right|^{2}\mathfrak{g}_{0}^{<}\left(t_{1},t\right)$,
$\mathfrak{g}_{0}\left(t_{1},t\right)$ is the unperturbed surface
Green's function of the linear chain, and $G_{0,A}\left(t,t_{1}\right)$
is the full surface Green's function of the dimers chain. Due to the
time periodicity, the expression for the current can be reduced to
matrix products in Floquet representation:
\begin{eqnarray}
J_{mn} & = & 2q\int_{-\infty}^{\infty}\sum_{k}\Re\left\{ G_{mk}^{r}\left(\omega\right)\Sigma_{kn}^{<}\left(\omega\right)\right.\nonumber \\
 &  & +\left.G_{mk}^{<}\left(\omega\right)\Sigma_{kn}^{a}\left(\omega\right)\right\} d\omega
\end{eqnarray}
where for simplicity we have included just the Floquet indices. In
this work we assume that the dimers chain is driven out of equilibrium
by the AC field, while the contact is in equilibrium at some chemical
potential $\mu_{L}$. In this case, all time independent terms become
diagonal in Floquet indices.

\begin{figure}
\includegraphics[scale=0.6]{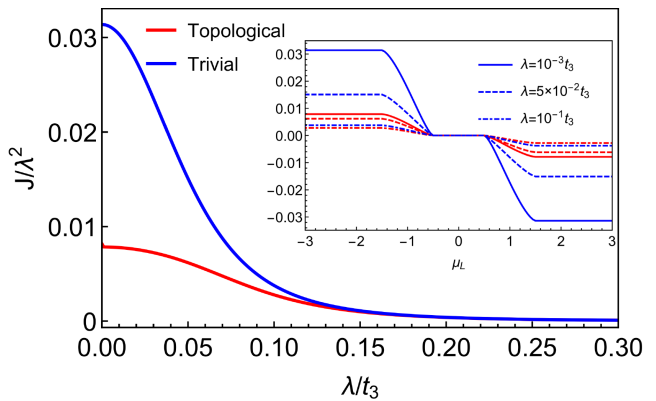}

\caption{\label{fig:StaticCurrent1}Average current vs coupling strength to
the contact, for the topological($t_{1}=0.5t_{2}$) and trivial($t_{1}=2t_{2}$)
phase (red and blue, respectively). For weak coupling we find a strong
suppression of the current in the topological phase, related with
the presence of a localized mode and a small S-DOS at site A (see
Fig.\ref{fig:DOS-EdgeDimers}). Increasing $\lambda$ supresses the
current in both cases, consequence of a decrease in the total S-DOS.
We have chosen $t_{3}=100$ in units of $t_{2}$. The inset shows
the I-V curve for different $\lambda$.}

\end{figure}

We first describe the I-V curves in the static case and the influence
of hybridization with the metallic contact. In Fig.\ref{fig:StaticCurrent1}
we plot the large bias current as a function of $\lambda$, for the
topological and the trivial phase. It shows that the presence of the
edge state lowers the average current, but it is still finite due
to a finite DOS at finite energy. We observe that the difference between
the two is especially large near the weak coupling limit $\lambda\ll t_{1},t_{2}$,
as they scale very differently around $\lambda=0$\cite{Benito-AC-Dimers}.
The main reason for this decrease is the strong localization of the
edge state, and the fact that it is the one that directly couples
the dimers chain to the metallic contact. The inset shows the I-V
curves for different values of $\lambda$, which show that all curves
collapse to a single one in the strong coupling limit. It is also
important that the contribution of the edge state to the current is
infinitesimally small, as the domain wall provides just one electron
to the current. Therefore, its presence contributes with an infinitesimal
change in the current at $\mu_{L}=0$. Furthermore, the broadening
of the mode due to hybridization does not seem to be captured in the
I-V plot, which makes its detection more difficult. Finally, we have
previously seen that the topological phase could be driven from the
trivial one by hybridization with the contact. Unfortunately, this
process is accompanied by a decrease in the S-DOS at the A site (now
highly hybridized with the contact), decreasing the current and making
more difficult the detection of the transition. Nevertheless we will
show below that the dependence of the current on the field amplitude
$\xi$, can help us to discriminate between the different phases.

\begin{figure}
\includegraphics[scale=0.6]{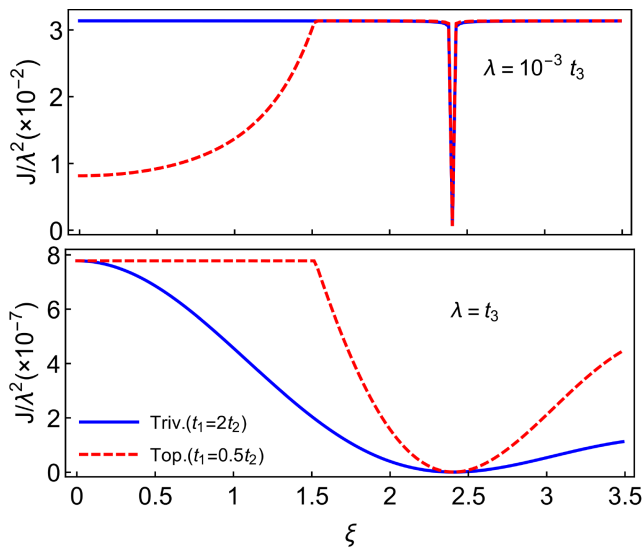}

\caption{\label{fig:Large-bias-current}Large bias current vs $\xi$, for weak
and strong (top and bottom, respectively) coupling to the metallic
contact. The current responds continuously to variations of the field
amplitude if the dimers chain has a localized energy mode, while the
trivial phase is insensitive. We have chosen $\Omega$ twice as large
as the undriven bandwidth and $t_{3}=100$, in units of $t_{2}$.}
\end{figure}

\begin{figure}
\includegraphics[scale=0.6]{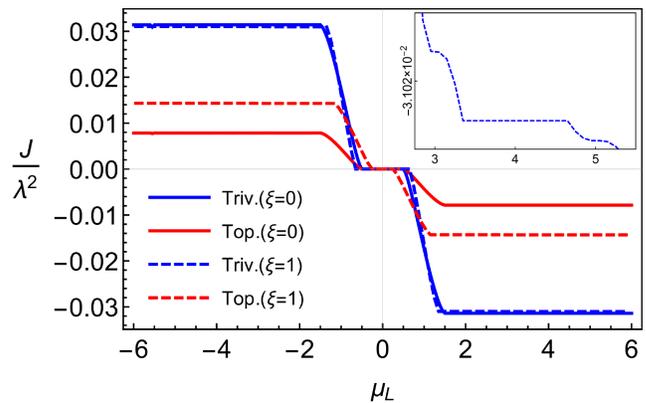}

\caption{\label{fig:Time-average-current}Time average current as a function
of the chemical potential $\mu_{L}$. The solid lines correspond to
the current in absence of driving, and the dashed lines correspond
to the average current for $\xi=1$ and $\Omega=4$. The inset shows
a zoom of the current when $\mu_{L}$ is measuring the contribution
from the photo-induced side-bands. We have chosen $\lambda=10^{-2}$
and $t_{3}=100$, in units of $t_{2}$.}
\end{figure}

We now focus on the non-equilibrium case and discuss the average current
and the I-V curve, as a function of the AC field parameters and for
different coupling strengths to the metallic contact. In Fig.\ref{fig:Large-bias-current}
we plot the time average current as a function of the field amplitude
$\xi$ for large bias and large $\Omega$. The top figure corresponds
to the weak coupling limit ($\lambda=10^{-3}t_{3}$), where the dimers
chain is slightly hybridized with the contact. It can be seen that,
in the trivial phase, increasing the field amplitude does not affect
the average current(blue solid line, for $t_{1}/t_{2}=2$), and it
remains constant until the first zero of $\mathcal{J}_{0}\left(\xi\right)$,
where the current suddenly drops to zero. This is equivalent to the
well known coherent destruction of tunneling mechanism\cite{CDT1,CDT4,CDT5,CDT2,CDT3},
where the dimers decouple. On the other hand, the topological phase
(red dashed line, for $t_{1}/t_{2}=0.5$) shows a continuous variation
of the current as a function of $\xi$ due to the presence of the
edge state. In this case, an increase of $\xi$ continuously reduces
the gap between the conduction and the valence band, and it is easier
for an electron localized in the edge state to jump to one of the
bulk bands, increasing the current. At $\xi\simeq1.5$ the two bands
close the gap, the phase becomes trivial, and then insensitive to
changes of $\xi$. The bottom figure plots the case of large hybridization
with the contact ($\lambda=t_{3}$), where as we previously discussed,
the role of the topological and trivial phase has been inverted, but
the current in the absence of the AC field could not distinguish between
the two of them. It turns out that as the AC field amplitude increases,
both cases show the same behavior as in the weak coupling case, ie.
the current in the presence of a localized state continuously changes
with $\xi$, while the current for the phase without an edge state
is locked until the bands close the gap again. This shows that the
$\xi$ dependence of the time averaged current can be used as a tool
to detect the presence of localized modes in the chain.
Finally, we analyze the I-V curves for different parameters of the
AC field; they are shown in Fig.\ref{fig:Time-average-current}. The
comparison between the equilibrium and out-of-equilibrium case (solid
and dashed, respectively) shows how the large bias current is unaffected
within the trivial phase (blue), while for the topological phase increases
to almost twice its original value; also, the renormalization of the
bands can be observed as an increase in the slope of the curves. The
inset shows that the current can capture the side-band structure previously
discussed, as the AC field induces new transport channels at multiples
of the driving frequency. In the high frequency regime, which is discussed
in this work, their contribution to the current is small; however,
their presence will increase as the system approaches resonance. It
is also important to notice that, as for the undriven case, the edge
states at $\omega=n\Omega$ are absent from the I-V curves due to
their small spectral weight.

\paragraph*{Conclusions:}

We have studied the non-equilibrium properties of a semi-infinite
dimers chain coupled to both, an AC electric field and to a metallic
contact at a different chemical potential. Combining Keldysh formalism
with the surface Green's functions method for semi-infinite systems
we have studied the thermodynamic limit, where finite size effects
do not affect the edge states or transport properties. Furthermore,
the formalism allows us to obtain results in the case of large hybridization
between the dimers chain and the metallic contact (strong coupling
limit). We find that with this method we can calculate to arbitrary
accuracy the surface Green's functions of the system, and therefore
analyze the fate of the edge states in a non-equilibrium topological
phase, and their contribution to the current. For the equilibrium
case we find that the strong coupling limit can alter the topological
properties of the dimers chain, as the hybridization with the metallic
contact can change the dimerization ground state of the system. We
find the characteristic current suppression and edge-state blockade
of a system with edge states for the case of small hybridization with
the contact\cite{Benito-AC-Dimers}; however, we have shown that as
the hybridization increases, this effect gets reduced, making more
difficult the distinction between the trivial and the topological
phase. In the presence of the high frequency AC field we have found
that the equilibrium edge states, which are initially localized in
one sub-lattice, gain spectral weight in the opposite one, and show
the appearance of side-bands, which contribute as extra transport
channels. Importantly, we have found that the average current shows
a different behavior, as a function of the AC field amplitude, when
the dimers chain has an edge state; in the trivial phase the current
is slightly affected by a change in the field amplitude, while the
topological phase shows a continuous variation of the average current
due to the hybridization between the edge state and the bulk bands.
Furthermore, this property seems to hold for large hybridization with
the metallic contact, which could be very helpful in a realistic situation. 

The study of intermediate frequencies would be interesting for future
works, as the topological properties change once different side-bands
cross. The adiabatic regime is also of interest, with the inclusion
of multi-frequency fields to simulate higher dimensional properties\cite{MultiFreq}.
Extensions of this work including domain wall dynamics would also
be interesting for applications in molecular electronics, where the
dynamics of solitons has been proposed to build molecular switches,
transistors and memories\cite{SolitonSwitch}. In this case, one could
take advantage of the external control provided by the AC field, and
combine the soliton dynamics with the external control of the coupling
to the electrons. Finally, although we have not discussed the effect
of dissipation and heating in the system, it can be important in periodically
driven systems and would require an analysis including a dissipative
bosonic and fermionic bath; however, this discussion is out of the
scope of the present manuscript and would require a more detailed
description of a concrete setup in order to include all the relevant
decoherence mechanisms\cite{Mitra1,PhysRevX.5.041050,Chamon1,Dissipation1}.

We would like to acknowledge P.C.E. Stamp, G. Platero, and M. Benito
for the critical reading of the manuscript. This work was supported
by NSER of Canada and MAT2014.

\bibliographystyle{phaip}
\bibliography{Keldysh&Topology}

\begin{widetext}

\appendix

\section{Surface Green's functions for semi-infinite chains}

In this section we calculate the surface Green's functions for the
linear and the dimers chain. We mention that the surface Green's functions
are interesting because they can be calculated exactly, and they provide
information about boundary states in systems with hardwall conditions.
For the calculation of the surface Green's functions we just need
to consider Dyson's equation and the recurrence relation obtained
in a semi-infinite system. Let us begin with the metallic contact,
modeled by a tight binding Hamiltonian:
\begin{equation}
H_{C}=\mu_{L}\sum_{i=0}^{\infty}n_{i}-t_{3}\sum_{\langle i,j\rangle}c_{i}^{\dagger}c_{j}+h.c.
\end{equation}
If we consider the case of the $x_{0}$ site initially decoupled from
the chain (i.e., $c_{0}^{\dagger}c_{1}$ and $c_{1}^{\dagger}c_{0}$
are removed from the Hamiltonian), and we re-attach it, the process
can be described with the Dyson's equation $\hat{\mathfrak{g}}=\hat{\mathfrak{g}}^{\left(0\right)}+\hat{\mathfrak{g}}^{\left(0\right)}\cdot\hat{V}\cdot\hat{\mathfrak{g}}$,
where $\hat{\mathfrak{g}}^{\left(0\right)}$ is the Green's function
for the two systems initially decoupled, $\hat{\mathfrak{g}}$ is
the Green's function for the system when they are coupled, and $\hat{V}$
is the hopping between the sites $0$ and $1$. In matrix form, the
Dyson's equation reads:
\begin{equation}
\left(\begin{array}{cc}
\mathfrak{g}_{0} & \mathfrak{g}_{01}\\
\mathfrak{g}_{10} & \mathfrak{g}_{1}
\end{array}\right)=\left(\begin{array}{cc}
\mathfrak{g}_{0}^{\left(0\right)} & 0\\
0 & \mathfrak{g}_{1}^{\left(0\right)}
\end{array}\right)+\left(\begin{array}{cc}
\mathfrak{g}_{0}^{\left(0\right)} & 0\\
0 & \mathfrak{g}_{1}^{\left(0\right)}
\end{array}\right)\left(\begin{array}{cc}
0 & -t_{3}\\
-t_{3} & 0
\end{array}\right)\left(\begin{array}{cc}
\mathfrak{g}_{0} & \mathfrak{g}_{01}\\
\mathfrak{g}_{10} & \mathfrak{g}_{1}
\end{array}\right)
\end{equation}
where $\mathfrak{g}_{1}^{\left(0\right)}$ corresponds to the unperturbed
surface Green's function at site $1$ of the semi-infinite chain.
The equation for $\mathfrak{g}_{0}$ results in:
\begin{align}
\mathfrak{g}_{0}= & \mathfrak{g}_{0}^{\left(0\right)}-\mathfrak{g}_{1}^{\left(0\right)}t_{3}\mathfrak{g}_{10}=\mathfrak{g}_{0}^{\left(0\right)}+t_{3}^{2}\mathfrak{g}_{0}^{\left(0\right)}\mathfrak{g}_{1}^{\left(0\right)}\mathfrak{g}_{0}
\end{align}
where we have used $\mathfrak{g}_{10}=-t_{3}\mathfrak{g}_{1}^{\left(0\right)}\mathfrak{g}_{0}$.
Finally, noticing that after attaching the site $0$ to the semi-infinite
chain, $\mathfrak{g}_{1}^{\left(0\right)}=\mathfrak{g}_{0}$ we obtain
a second order equation with solution:
\begin{equation}
\mathfrak{g}_{0}\left(\omega\right)=\frac{1\pm\sqrt{1-4t_{3}^{2}\left[\mathfrak{g}_{0}^{\left(0\right)}\left(\omega\right)\right]^{2}}}{2t_{3}^{2}\mathfrak{g}_{0}^{\left(0\right)}\left(\omega\right)}
\end{equation}
Finally, we use the single site Green's function $\mathfrak{g}_{0}^{\left(0\right)}=\left(\omega-\mu_{L}\right)^{-1}$
and find:
\begin{equation}
\mathfrak{g}_{0}\left(\omega\right)=\frac{\omega-\mu_{L}}{2t_{3}^{2}}\left[1\pm\sqrt{1-\left(\frac{2t_{3}}{\omega-\mu_{L}}\right)^{2}}\right]
\end{equation}
where we can fix the minus sign in order to obtain the right density
of states:
\begin{equation}
\rho\left(\omega\right)=-\frac{1}{\pi}\lim_{\epsilon\rightarrow0^{+}}\Im\left\{ \mathfrak{g}_{0}\left(\omega+i\epsilon\right)\right\} 
\end{equation}
which is non-vanishing when $\left|\omega-\mu_{L}\right|<2t_{3}$.
For the case of a dimers chain one can proceed in a similar way, this
time including a sub-lattice degree of freedom.

\section{Floquet Green's functions}

Here we calculate the Floquet-Green's function for the dimers chain
coupled to an AC field. We define the dimers chain Green's function
and its Wigner form as ($t_{+}=\left(t+t^{\prime}\right)/2$ and $t_{-}=t-t^{\prime}$):
\begin{align}
G_{k}^{\mu,\nu}\left(t,t^{\prime}\right) & =-i\theta\left(t-t^{\prime}\right)\langle\left\{ d_{k,\mu}\left(t\right),d_{k,\nu}^{\dagger}\left(t^{\prime}\right)\right\} \rangle\\
G\left(t_{+},t_{-}\right) & =\frac{1}{2\pi}\int d\omega e^{-i\omega t_{-}}\sum_{n}e^{-in\Omega t_{+}}G_{n}\left(\omega\right)\\
G_{n}\left(\omega\right) & =\int e^{i\omega t_{-}}dt_{-}\frac{1}{T}\int_{0}^{T}e^{in\Omega t_{+}}G\left(t_{+},t_{-}\right)dt_{+}
\end{align}
which corresponds to the propagation of a fermion with momentum $k$
under the Hamiltonian:
\begin{eqnarray}
\tilde{H}_{D}\left(t\right) & = & \mu_{D}\sum_{\alpha,i}d_{i,\alpha}^{\dagger}d_{i,\alpha}-\sum_{i,j,\alpha,\beta}\tilde{t}_{i,j}^{\alpha,\beta}\left(t\right)d_{i,\alpha}^{\dagger}d_{j,\beta}=\mu_{D}\sum_{\alpha,\mathbf{k}}n_{\mathbf{k},\alpha}-\sum_{k,\alpha,\beta}\hat{t}_{k}^{\alpha,\beta}\left(t\right)d_{k,\alpha}^{\dagger}d_{k,\beta}
\end{eqnarray}
where the time dependent hoppings are $\tilde{t}_{i,j}^{\alpha,\beta}\left(t\right)=t_{i,j}^{\alpha,\beta}e^{i\xi\left(x_{i,\alpha}-x_{j,\beta}\right)\sin\left(\Omega t\right)}$
and $\xi=\frac{qV_{0}}{\Omega}$, and concretely in our case:
\begin{equation}
\hat{t}_{k}^{\alpha,\beta}\left(t\right)=\left(\begin{array}{cc}
0 & t_{1}+t_{2}e^{i\left(k+\xi\sin\left(\Omega t\right)\right)}\\
t_{1}+t_{2}e^{-i\left(k+\xi\sin\left(\Omega t\right)\right)} & 0
\end{array}\right)=\left(\begin{array}{cc}
0 & t_{1}+t_{2}e^{ik}\sum_{l}J_{l}\left(\xi\right)e^{il\Omega t}\\
t_{1}+t_{2}e^{-ik}\sum_{l}J_{-l}\left(\xi\right)e^{il\Omega t} & 0
\end{array}\right)
\end{equation}
The equation of motion for the Green's function is given by:
\begin{eqnarray}
\left(\frac{i}{2}\partial_{t_{+}}+i\partial_{t_{-}}-\mu_{D}\right)G_{k}^{\mu,\nu}\left(t_{+},t_{-}\right) & = & \delta_{\mu,\nu}\delta\left(t_{-}\right)-\hat{t}_{k}^{\mu,-\mu}\left(t\right)G_{k}^{-\mu,\nu}\left(t_{+},t_{-}\right)
\end{eqnarray}
and Fourier transforming to frequency space we finally get ($\mu,\nu=A,B$):
\begin{eqnarray}
\left(n\frac{\Omega}{2}+\omega-\mu_{D}\right)G_{k;n}^{\mu,\nu}\left(\omega\right) & = & \delta_{n,0}\delta_{\mu,\nu}-t_{1}G_{k;n}^{-\mu,\nu}\left(\omega\right)-t_{2}\sum_{l}\mathcal{J}_{\mu l}\left(\xi\right)e^{i\mu k}G_{k;n+l}^{-\mu,\nu}\left(\omega+l\frac{\Omega}{2}\right)
\end{eqnarray}
Therefore, the equation for the $n=0$ Green's function is:
\begin{eqnarray}
\left(\omega-\mu_{D}\right)G_{k;0}^{\mu,\nu}\left(\omega\right) & = & \delta_{\mu,\nu}-t_{1}G_{k;0}^{-\mu,\nu}\left(\omega\right)-t_{2}\sum_{l}\mathcal{J}_{\mu l}\left(\xi\right)e^{i\mu k}G_{k;l}^{-\mu,\nu}\left(\omega+l\frac{\Omega}{2}\right)
\end{eqnarray}
and the general equation of motion, for all the Green's functions
included in the calculation, is:
\begin{eqnarray}
\left(\omega+l\Omega-\mu_{D}\right)G_{k;l}^{\mu,\nu}\left(\omega+l\frac{\Omega}{2}\right) & = & \delta_{l,0}\delta_{\mu,\nu}-t_{1}G_{k;l}^{-\mu,\nu}\left(\omega+l\frac{\Omega}{2}\right)-t_{2}\sum_{r}\mathcal{J}_{\mu r}\left(\xi\right)e^{i\mu k}G_{k;l+r}^{-\mu,\nu}\left(\omega+\left(l+r\right)\frac{\Omega}{2}\right)
\end{eqnarray}
This system of equations can be solved to arbitrary accuracy and mapped
into the Floquet representation. The calculation of the AC driven
surface Green's functions is similar, but we need to consider the
Dyson's equation to represent the process of coupling the single site
to the semi-infinite chain, and this time, a self-energy which is
time dependent. The Dyson's equation is given by:
\begin{equation}
\hat{G}\left(t,t^{\prime}\right)=\hat{g}\left(t,t^{\prime}\right)+\int dt_{1}\int dt_{2}\hat{g}\left(t,t_{1}\right)\hat{\Sigma}\left(t_{1},t_{2}\right)\hat{G}\left(t_{2},t^{\prime}\right)
\end{equation}
where:
\begin{eqnarray}
\hat{G}\left(t,t^{\prime}\right) & = & \left(\begin{array}{cc}
\hat{G}_{0}\left(t,t^{\prime}\right) & \hat{G}_{0C}\left(t,t^{\prime}\right)\\
\hat{G}_{C0}\left(t,t^{\prime}\right) & \hat{G}_{C}\left(t,t^{\prime}\right)
\end{array}\right)\\
\hat{g}\left(t,t^{\prime}\right) & = & \left(\begin{array}{cc}
\hat{g}_{0}\left(t,t^{\prime}\right) & 0\\
0 & \hat{G}_{0}\left(t,t^{\prime}\right)
\end{array}\right)\\
\hat{\Sigma}\left(t,t^{\prime}\right) & = & \left(\begin{array}{cc}
0 & \tilde{t}\left(t\right)\\
\tilde{t}\left(t\right)^{\ast} & 0
\end{array}\right)\delta\left(t-t^{\prime}\right)
\end{eqnarray}
and we have defined $\tilde{t}\left(t\right)=\tilde{t}e^{i\xi\sin\left(\Omega t\right)}$
and $\tilde{t}$ is a matrix in the sub-lattice indices for the case
of a dimers chain. In order to deal with this integral equation it
is useful to go to the Floquet representation:
\begin{align}
\hat{G}_{mn}\left(\omega\right) & \equiv\hat{G}_{m-n}\left(\omega+\frac{m+n}{2}\Omega\right)\\
\hat{g}_{mn}\left(\omega\right) & \equiv\hat{g}_{m-n}\left(\omega+\frac{m+n}{2}\Omega\right)\\
\hat{\Sigma}_{mn}\left(\omega\right) & \equiv\hat{\Sigma}_{m-n}\left(\omega+\frac{m+n}{2}\Omega\right)
\end{align}
where the Dyson's equation transforms into a matrix multiplication:
\begin{equation}
\hat{G}_{mn}\left(\omega\right)=\hat{g}_{mn}\left(\omega\right)+\sum_{m^{\prime}n^{\prime}}\hat{g}_{mm^{\prime}}\left(\omega\right)\hat{\Sigma}_{m^{\prime}n^{\prime}}\left(\omega\right)\hat{G}_{n^{\prime}n}\left(\omega\right)
\end{equation}
For the explicit calculation of the dimers chain we need to calculate
the Wigner transform of the self-energy, which leads to the next Floquet
representation:
\begin{align}
\hat{\Sigma}_{mn} & =-t_{2}\left(\begin{array}{cccc}
0 & 0 & 0 & 0\\
0 & 0 & \mathcal{J}_{n-m}\left(\xi\right) & 0\\
0 & \mathcal{J}_{m-n}\left(\xi\right) & 0 & 0\\
0 & 0 & 0 & 0
\end{array}\right)
\end{align}
With these expressions one can solve the Dyson's equation for the
surface Green's function in presence of driving.

\section{Current calculation}

In this Appendix we write the explicit expressions for the time dependent
current through the hybrid system. For that we make use of the Keldysh
formalism, which allows to treat the problem in a very simple and
systematic way. We start by separating the total Hamiltonian into
its central system, reservoir and tunneling parts:
\begin{equation}
H=H_{D}+H_{C}+H_{T}
\end{equation}
For the current calculation we just need to focus in the tunneling
Hamiltonian:
\begin{equation}
H_{T}=\sum_{i,j,\sigma}\left(\lambda_{j;\sigma,i}^{\ast}\left(t\right)c_{j}^{\dagger}d_{\sigma,i}+\lambda_{j;\sigma,i}\left(t\right)d_{\sigma,i}^{\dagger}c_{j}\right)
\end{equation}
as the metallic contact and dimers chain Hamiltonians can be kept
quite general for this discussion. To determine the current operator
we begin by calculating the time variation of the number of particles
in the reservoir $\hat{Q}=q\sum_{i}c_{i}^{\dagger}c_{i}$ ($\hbar=1$):
\begin{align}
\hat{J}\left(t\right) & =\partial_{t}\hat{Q}=iq\sum_{j\in R}\sum_{\sigma,i\in S}\left\{ \lambda_{j;\sigma,i}\left(t\right)d_{i,\sigma}^{\dagger}c_{j}-\lambda_{j;\sigma,i}^{\ast}\left(t\right)c_{j}^{\dagger}d_{i,\sigma}\right\} 
\end{align}
Therefore, the average current is related with the lesser Green's
functions:
\begin{align}
G_{j,\alpha;l,\sigma}^{<}\left(t,t^{\prime}\right) & =i\langle d_{l,\sigma}^{\dagger}\left(t^{\prime}\right)c_{j,\alpha}\left(t\right)\rangle\\
G_{l,\sigma;j,\alpha}^{<}\left(t,t^{\prime}\right) & =i\langle c_{j,\alpha}^{\dagger}\left(t^{\prime}\right)d_{l,\sigma}^{\dagger}\left(t\right)\rangle
\end{align}
and can be rewritten as:
\begin{equation}
J\left(t\right)=2q\sum_{i,j,\sigma}\Re\left\{ \lambda_{j;\sigma,i}\left(t\right)G_{j;i,\sigma}^{<}\left(t,t\right)\right\} 
\end{equation}
In conclusion, the time dependent current requires to calculate the
equal time, lesser Green's function $G_{j;i,\sigma}^{<}\left(t,t\right)$
only. Furthermore, as we will assume that the tunnel Hamiltonian connects
the two sites at the end of each chain, the expression will only depend
on the surface Green's functions. Using the equation of motion technique
and the Langreth rules for the contour ordered Green's function, one
can write the expression for the lesser Green's function as:
\begin{equation}
G_{j;i,\sigma}^{<}\left(t,t\right)=\sum_{i^{\prime},\sigma^{\prime}}\int dt_{1}\lambda_{j;i^{\prime},\sigma^{\prime}}^{\ast}\left(t_{1}\right)\left[G_{i,\sigma;i^{\prime},\sigma^{\prime}}^{r}\left(t,t_{1}\right)\mathfrak{g}_{j}^{<}\left(t_{1},t\right)+G_{i,\sigma;i^{\prime},\sigma^{\prime}}^{<}\left(t,t_{1}\right)\mathfrak{g}_{j}^{a}\left(t_{1},t\right)\right]
\end{equation}
were we have separated the mixed Green's function into a product of
surface Green's functions, and $\mathfrak{g}$ corresponds to the
unperturbed surface Green's function of the metallic contact. The
expression for the current becomes:
\begin{equation}
J\left(t\right)=2q\sum_{i,j,\sigma}\sum_{i^{\prime},\sigma^{\prime}}\int_{-\infty}^{\infty}dt_{1}\Re\left\{ \lambda_{j;\sigma,i}\left(t\right)\lambda_{j;i^{\prime},\sigma^{\prime}}^{\ast}\left(t_{1}\right)\left[G_{i,\sigma;i^{\prime},\sigma^{\prime}}^{r}\left(t,t_{1}\right)\mathfrak{g}_{j}^{<}\left(t_{1},t\right)+G_{i,\sigma;i^{\prime},\sigma^{\prime}}^{<}\left(t,t_{1}\right)\mathfrak{g}_{j}^{a}\left(t_{1},t\right)\right]\right\} 
\end{equation}
and identifying the self-energies:
\begin{align}
\Sigma_{i,\sigma;i^{\prime},\sigma^{\prime}}^{a}\left(t_{1},t\right) & =\sum_{j}\lambda_{j;\sigma,i}\left(t\right)\lambda_{j;i^{\prime},\sigma^{\prime}}^{\ast}\left(t_{1}\right)\mathfrak{g}_{j}^{a}\left(t_{1},t\right)\\
\Sigma_{i,\sigma;i^{\prime},\sigma^{\prime}}^{<}\left(t_{1},t\right) & =\sum_{j}\lambda_{j;\sigma,i}\left(t\right)\lambda_{j;i^{\prime},\sigma^{\prime}}^{\ast}\left(t_{1}\right)\mathfrak{g}_{j}^{<}\left(t_{1},t\right)
\end{align}
we obtain:
\begin{equation}
J\left(t\right)=2q\sum_{j,i,\sigma}\sum_{i^{\prime},\sigma^{\prime}}\int_{-\infty}^{\infty}\Re\left\{ G_{i,\sigma;i^{\prime},\sigma^{\prime}}^{r}\left(t,t_{1}\right)\Sigma_{i,\sigma;i^{\prime},\sigma^{\prime}}^{<}\left(t_{1},t\right)+G_{i,\sigma;i^{\prime},\sigma^{\prime}}^{<}\left(t,t_{1}\right)\Sigma_{i,\sigma;i^{\prime},\sigma^{\prime}}^{a}\left(t_{1},t\right)\right\} dt_{1}
\end{equation}
The problem of calculating the current for a time dependent system
has been reduced to the calculation of the self-energies and Green's
functions for the dimers chain. The form of the previous expression
allows for one further simplification when dealing with AC fields.
We can consider the Floquet-Keldysh formalism and reduce the time
integrals to matrix multiplications. For that purpose let us generalize
the current operator to a two-2 time function:
\begin{equation}
J\left(t,t^{\prime}\right)=2q\sum_{i,j,\sigma}\sum_{i^{\prime},\sigma^{\prime}}\int_{-\infty}^{\infty}\Re\left\{ G_{i,\sigma;i^{\prime},\sigma^{\prime}}^{r}\left(t,t_{1}\right)\Sigma_{i,\sigma;i^{\prime},\sigma^{\prime}}^{<}\left(t_{1},t^{\prime}\right)+G_{i,\sigma;i^{\prime},\sigma^{\prime}}^{<}\left(t,t_{1}\right)\Sigma_{i,\sigma;i^{\prime},\sigma^{\prime}}^{a}\left(t_{1},t^{\prime}\right)\right\} dt_{1}
\end{equation}
which can be rewritten in Wigner coordinates as:
\begin{eqnarray}
J\left(t_{+},t_{-}\right) & = & \sum_{l}\int_{-\infty}^{\infty}J_{l}\left(\omega\right)e^{-il\Omega t_{+}-i\omega t_{-}}\frac{d\omega}{2\pi}\\
J_{l}\left(\omega\right) & = & \frac{1}{T}\int_{0}^{T}dt_{+}\int_{-\infty}^{\infty}J\left(t_{+},t_{-}\right)e^{il\Omega t_{+}+i\omega t_{-}}dt_{-}
\end{eqnarray}
Therefore, the equal time current is given by:
\begin{eqnarray}
J\left(t\right) & = & \sum_{l}e^{-il\Omega t}\int_{-\infty}^{\infty}J_{l}\left(\omega\right)\frac{d\omega}{2\pi}=\sum_{l}e^{-il\Omega t}J_{l}
\end{eqnarray}
obtained by the inverse Fourier transformed of $J_{l}\left(\omega\right)$
integrated over all frequencies $\omega$. Therefore, we just need
to calculate the Floquet matrix (in the next expression we omit the
site and sub-lattice indices for simplicity):
\begin{equation}
J_{mn}\left(\omega\right)=2q\sum_{k}\Re\left\{ G_{mk}^{r}\left(\omega\right)\Sigma_{kn}^{<}\left(\omega\right)+G_{mk}^{<}\left(\omega\right)\Sigma_{kn}^{a}\left(\omega\right)\right\} 
\end{equation}
and integrate over all $\omega$ in order to obtain the different
Fourier components of the current. Concretely for the constant contribution
we just need to obtain the diagonal terms $m=n$.

As in our model we assume that the contact is in equilibrium, and
that the tunneling $\lambda$ is time independent, the expressions
for the self energies and the surface Green's functions for the linear
chain highly simplify:
\begin{eqnarray}
\mathfrak{g}_{0}^{<}\left(\omega\right) & = & if\left(\omega\right)A_{0}\left(\omega\right)\\
\Sigma_{0,A}^{a}\left(\omega\right) & = & \left|\lambda\right|^{2}\mathfrak{g}_{0}^{a}\left(\omega\right)\\
\Sigma_{0,A}^{<}\left(\omega\right) & = & \left|\lambda\right|^{2}\mathfrak{g}_{0}^{<}\left(\omega\right)
\end{eqnarray}
where $f\left(\omega\right)=\left(e^{\beta\omega}+1\right)^{-1}$
is the Fermi function. Their representation in Floquet form is trivial,
as it corresponds to a diagonal form. Therefore, for the calculation
one just needs to include the full surface Green's functions in Floquet
form, obtained in the main text.

\end{widetext}
\end{document}